\documentclass[%
 reprint,
 showkeys,
nofootinbib,
nobibnotes,
 amsmath,amssymb,
 aps,
floatfix,
]{revtex4-2}

\usepackage{graphicx}
\usepackage{dcolumn}
\usepackage{bm}

\begin{document}

\preprint{APS/123-QED}

\title{Photodetachment dynamics using nonlocal dicrete-state-in-continuum model}

\author{Martin \v{C}\'{\i}\v{z}ek}%
 \email{Martin.Cizek@mff.cuni.cz}
\affiliation{Charles University, Faculty of Mathematics and Physics, Institute of Theoretical Physics, \\
             V Hole\v{s}ovi\v{c}k\'{a}ch 2, 180 00 Prague, Czech Republic}


\date{\today}

\begin{abstract}
 In this preprint I propose that the non local discrete-state-in-continuum model previously 
 successfully used to describe the inelastic electron molecule collisions can also be used 
 to model the electron photo-detachment from the molecular anions. The basic theory is 
 sketched and the approach is tested on the model of electron photodetachment 
 from diatomic molecular anion. 
\end{abstract}

\maketitle


\section{\label{sec:intro}Introduction} 


The nonlocal discrete-state-in-continuum model is very successful approach in the description 
of the low-energy inelastic electron collisions \cite{D1991,Kniha_2012}
leading to vibrational excitation (VE)
\begin{equation}
 \label{eq:VE}
 e^- + M(v_i) \to M^- \to e^- + M(v_f)
\end{equation}
and the dissociative attachment (DA)
\begin{equation}
 \label{eq:DA}
 e^- + M(v_i) \to M^- \to A^- + B.
\end{equation}
The key ingredient of the theory is that both processes proceed through formation of a metastable
molecular anion state $M^-$ out of equilibrium, that undergoes the vibronic dynamics and 
decays either back into electron-molecule scattering continuum $e^- + M(v_f)$ 
or dissociates into fragment $A^- + B$. The convenient method to study the dynamics 
of such process is through electron energy losss spectroscopy (EELS) \cite{A1989}.
This technique is based on the energy conservation 
\begin{equation}\label{eq:Econserv}
  E = \epsilon_i + E_{v_i} = \epsilon_f + E_{v_f}, 
\end{equation}
where $\epsilon$ are electron energies and $E_v$ energies of vibrational states
of the molecule for the initial $|\nu_i\rangle$ and the final $|\nu_f\rangle$
vibrational states, before and after the collision.
Furthermore the cross section of the processes is enhanced if the total energy $E$ 
attains value close to energies of the metastable vibronic states of the temporary 
anion $M^-$. The most complete experimental picture is provided by scanning through 
both initial $\epsilon_i$ and final $\epsilon_f$ electron energies (or equivalently 
energy loss $\Delta\epsilon=\epsilon_i-\epsilon_f$) creating thus 2D EELS picture. 
The dependence on the scattering angle for electron can also be monitored.
Such spectra are still not well understood \cite{RA2013,RA2015,AMN2020,UNP_Allan}.

The nonlocal discrete state in continuum theory has recently been successfully used 
to calculate the 2D EELS for CO$_2$ molecule \cite{the_letter,partI,partII}. In this 
paper we propose to use the same kind of theory to calculate the electron 
spectrum for the photodetachment of an electron $e^-$
from a molecular anion $M^-$ 
\begin{equation}
 \label{eq:photod}
 \gamma + M^- \to (M^-)^* \to e^- + M(v_f),
\end{equation}
with initial energy $E$ of the system determined by the energy of the photon $\gamma$
shone on the anion to excite it to the state $(M^-)^*$. The vibronic 
dynamics of this moleculer metastable anion state is driven by the same principles 
as in the case of electron-molecule collisions. The energies of the released 
electron can be monitored as function of the photon energy giving thus 
the 2D spectrum similar to the 2D EELS \cite{AMN2020,RNA2022}. 

The modeling of the 2D photodetachment spectrum can thus proceed along the same 
lines as for 2D EELS and we can use the iterative methods recently developed 
to threat the dynamics for polyatomic molecules \cite{Bcl_Martina,Ms_Martina,Phd_Dvorak} 
also for the electron photodetachment.

In this paper we develop the theory of the resonance inelastic photodetachment 
process and propose to treat the resulting equations numerically with the codes 
developed for the electron-molecule collisions. The theory also includes the 
photodissociation process in analogy with the dissociative attachment (\ref{eq:DA}).
We also propose a simple model inspired by LiH$^-$ \cite{CDH2018,Ms_Dvorak}
to test the numerical methods and to discuss the resulting phenomena.

\section{\label{sec:theory}Theory}

\begin{figure}[th]
\centerline{\includegraphics[width=0.49\textwidth]{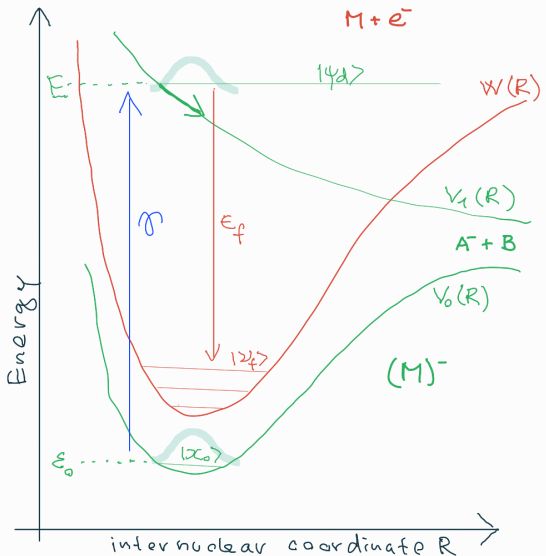}}
\caption{\label{fig:SchemPot}%
         Schematic picture of the photodetachment process. All potentials, energies and states related to anion 
         are drawn in green. The neutral molecule and the continuum-electron related quantities are in red and 
         and the absorbed photon energy in blue.}
\end{figure}

In this section we explain the basic ideas used to theoretically treat the inelastic 
resonance photodetachment based on the projection formalism for description of the 
dynamics of the discrete electron state in electron scattering continuum. The photon 
absoption is treated in the dipole approximation, but we do not see obstacles to include 
also higher order corections constidering the photon absorption. Once the photon is absorbed 
and the metastable negative ion is formed, the dynamics is treated in complete analogy 
with electron-molecule collisions where the anion is formed by electron attachment
(see Fig.~\ref{fig:SchemPot} for the schematic scatch of the process). 
The following paragraph thus describes the dynamics in similar way as in electron-molecule 
collisions (we refer to \cite{D1991,Kniha_2012} for reviews on the approach).

\subsection{Photodetachment in dipole approximation}

The initial state $|\Psi_0\rangle\rangle$ for the photodetachment process is the assumed 
to be the ground state of the molecular anion $M^-$, which we consider in Born-Oppenheimer 
approximation. We used the double-ket notation to stress that the wavefunction is product
of vibrational and electronic part $|\Psi_0\rangle\rangle=|\chi_0\rangle|\phi_0\rangle$, where 
$|\phi_0\rangle$ is the ground electronic state of the anion and the vibrational wavefunction 
$\chi_0(R)$ solves the usual Schr\"{o}dinger equation
\begin{equation}\label{eq:chi0}
  \left[T_N+V_0(R)\right] \chi_0(R) = \varepsilon_0 \chi_0(R),
\end{equation}
with the potential energy surface $V_0(R)$ depending on the nuclear geometry $R$, $T_N$
is the vibrational kinetic energy operator and $\varepsilon_0$ is energy of the initial 
vibrational state $|\chi_0\rangle$. The main goal of this paper is to evaluate the photodetachment 
amplitude  
\begin{equation}\label{eq:A}
  A = \langle\langle\Psi^{(-)}|D|\Psi_0\rangle\rangle
    = \langle\langle\Psi^{(-)}|D|\chi_0\rangle|\phi_0\rangle,
\end{equation}
where $D$ is the electrical dipole operator and $|\Psi^{(-)}\rangle\rangle$
is the scattering wavefunction subjected to outgoing boundary condition fixing 
the final state vibrational state of the neutral molecule $|\nu_f\rangle$ and 
the outgoing electron state $|\epsilon_f\rangle$, with the energy subjected to 
conservation law
\begin{equation}\label{eq:EconservP}
  E = \varepsilon_{\gamma} + \varepsilon_0 = \epsilon_f + E_{v_f}, 
\end{equation}
with photon energy $\varepsilon_{\gamma}$ and vibrational energy 
of the final state of the molecule $E_{v_f}$. 

We apply the discrete-state-in-continuum model and the projection-operator formalism 
to calculate the scattering wavefunction $|\Psi^{(-)}\rangle\rangle$. The starting 
point is the assumption of the existence of the diabatic basis in the Hilbert space 
of electrons with the fixed nuclei of the molecule, consisting of 1) at least two 
discrete states: already described ground state of the anion $|\phi_0\rangle$
and the excited metastable anion state $|\phi_1\rangle$ and 2) electron 
scatering continuum states. 
More discrete states can in principle be included but here we will limit the discussion 
to one bound and one discrete metastable state for simplicity. We can define 
the projector to the discrete state part of the electronic Hilbert space
\begin{equation}
    \mathcal{Q} = \sum_d \vert\phi_d\rangle\langle\phi_d\vert.
\end{equation} 
and the complementary operator
\begin{equation}
    \mathcal{P}=I - \mathcal{Q}
\end{equation}
projecting on the background electron continuum $e^- + M$.
The basis in this subspace can be chosen as solutions the background scattering problem 
\begin{equation}\label{eq:bckgr}
  \mathcal{P}\mathcal{H}_{el}\mathcal{P}\vert\Phi_0,\epsilon\mu\rangle
  =(W+\epsilon)\vert\Phi_0,\epsilon\mu\rangle,
\end{equation}
$W(R)$ is the potential energy surface of the neutral molecule, 
i.\ e.\ the energy of the ground electronic state $|\Phi_0\rangle$ of the neutral molecule 
and energy $\epsilon$ and quantum number $\mu$ identify the state of outgoing 
electron. Here we consider only the case that the neutral molecule has only 
one energetically accessible electronic state and we will suppress the symbol
$\Phi_0$ in the notation. We also assume only one dominant partial wave and suppress 
also the symbol $\mu$. We thus ended with the basis states $|\epsilon\rangle$
which can be used to expand the projector on the continuum part of the Hilbert 
stat for each $R$
\begin{equation}
  \mathcal{P}=\int |\epsilon\rangle\langle\epsilon| d\epsilon.
\end{equation}
Now the fixed nuclei electronic hamiltonian $\mathcal{H}_{el}$ can be expandent in the 
basis
\begin{eqnarray}
  \langle\phi_d|\mathcal{H}_{el}|\phi_{d'}\rangle    &=& V_d(R)\delta_{dd'},\\
  \langle\phi_d|\mathcal{H}_{el}|\epsilon\rangle     &=& V_{d\epsilon}(R),\\
  \langle\epsilon|\mathcal{H}_{el}|\epsilon'\rangle  &=& [W_0(R)+\epsilon]\,\delta(\epsilon-\epsilon').
\end{eqnarray}
Note that we neglected the coupling 
$\langle\phi_0|\mathcal{H}_{el}|\phi_{1}\rangle \simeq 0$ 
of the two discrete states, which assumes well isolated bound state $|\phi_{0}\rangle$
with noncrossing potentials $V_0(R)$ and $V_1(R)$. This asumption can be released but we will avoid it
in this work. We will also neglect the coupling of the bound state to the continuum by setting 
$V_{d\epsilon}(R)=0$ for $d=0$, but we include the nonzero amplitude $V_{1\epsilon}(R)$ which is 
responsible for the electron autodetachment from the state $|\phi_1\rangle$. 

The wavefunction $|\Psi^{(-)}\rangle\rangle$ can also be expanded in this basis
\begin{equation}\label{eq:PsiExp}
  |\Psi^{(-)}\rangle\rangle = 
    |\psi_{d}\rangle |\phi_d\rangle + \int |\psi_{\epsilon}\rangle |\epsilon\rangle d\epsilon.
\end{equation}
Note that due to the decoupling of the ground state $d=0$ we can consider only $d=1$ in this expansion.
This wavefunction is subjected to the same outgoing boundary condition like in the case 
of electron-molecule scattering and the $R$-depenent expansion coefficients $\psi_d(R)\equiv|\psi_d\rangle$ 
and $\psi_{\epsilon}(R)\equiv|\psi_{\epsilon}\rangle$ can be found in the same wave like 
in that case \cite{D1991,Kniha_2012}. In the case of the vibrational excitation process 
the relevant T-matrix element reads
(\ref{eq:VE})
\begin{equation}\label{eq:TVE}
  T_{VE} = \langle\langle\Psi^{(-)}|V|\nu_i\rangle|\epsilon_i\rangle
         = \langle\psi_d|V_{d\epsilon_i}|\nu_i\rangle,
\end{equation}
where $V=\mathcal{P}\mathcal{H}_{el}\mathcal{Q}+\mathcal{Q}\mathcal{H}_{el}\mathcal{P}$.
Notice that the last expression uses only the $\psi_d$ component of the expansion 
(\ref{eq:PsiExp}). Now we will remind how this component is evaluated and we 
would also like to process the expression (\ref{eq:A}) for the photodetachment 
amplitude in analogy with the expression (\ref{eq:TVE}) for the vibrational excitation 
process. The components of the wavefunction (\ref{eq:PsiExp}) can be found by solving 
the Schr\"{o}dinger equation with the hamiltonian $H=T_N+\mathcal{H}_{el}$ with 
the appropriate boundary condition 
\begin{eqnarray}
  |\psi_d\rangle & = & 0 
     + [E-H_d]^{-1} \int V_{d\epsilon} |\psi_{\epsilon}\rangle d\epsilon,
\\
  |\psi_{\epsilon}\rangle & = & |\nu_f\rangle \delta(\epsilon-\epsilon_f) 
     + [E-h_0-\epsilon]^{-1} V_{d\epsilon}|\psi_d\rangle.
\end{eqnarray}
By substituting the second of the equations into the first and slight rearangement 
we get
\begin{equation}\label{eq:SE_d}
  \left[E-H_d-F^{\dag}(E-h_0)\right]|\psi_d\rangle = V_{d\epsilon_f}|\nu_f\rangle,
\end{equation}
where 
\begin{eqnarray}
  H_d &=& T_N + V_d(R),\\
  h_0 &=& T_N + W(R),\\ \label{eq:Fnonloc}
  F(\varepsilon) &=& 
        \int_0^\infty V_{d\epsilon}(R)[\varepsilon-\epsilon+i\eta]^{-1} V_{d\epsilon}(R') d\epsilon.
\end{eqnarray}
Note that the final vibrational states of the neutral molecule after photodetachment are 
solution of the Schr\"odinger equation
\begin{equation}
    h_0 |\nu_f\rangle = E_{\nu_f}|\nu_f\rangle.
\end{equation}
The equation (\ref{eq:SE_d}) is used in theory of VE process to solve the dynamics numerically.
The formal solution can be written as 
\begin{equation}\label{eq:psi_d}
  |\psi_d\rangle = \left[E-H_d-F^{\dag}(E-h_0)\right]^{-1} V_{d\epsilon_f}|\nu_f\rangle .
\end{equation}
This leads to the well known simple expression for the vibrational excitation 
\begin{equation}\label{eq:TVE2}
  T_{VE} = \langle\nu_f|V_{d\epsilon_f} \left[E-H_d-F(E-h_0)\right]^{-1}  V_{d\epsilon_i}|\nu_i\rangle.
\end{equation}
For the photodetachment amplitude we also need the continuum component
{\small
\begin{eqnarray}\label{eq:psi_e}
  |\psi_{\epsilon}\rangle & = & 
      |\nu_f\rangle\delta(\epsilon-\epsilon_f) + \\ 
    &+& \left[E-h_0-\epsilon\right]^{-1} V_{d\epsilon}\left[E-H_d-F^{\dag}(E-h_0)\right]^{-1} 
       V_{d\epsilon_f}|\nu_f\rangle \nonumber.
\end{eqnarray}
}
Before we substitute the solution (\ref{eq:PsiExp}) with the components (\ref{eq:psi_d}) 
and (\ref{eq:psi_e}) into photodetachment amplitude (\ref{eq:A}) we define the fixed-$R$
transition dipole moments to discrete state $|\phi_1\rangle$ and to background continuum 
$|\epsilon\rangle$
\begin{eqnarray}
  \mu_d(R) & = & \langle\phi_1|D|\phi_0\rangle, \\
  \mu_{\epsilon}(R) & = & \langle\epsilon|D|\phi_0\rangle
\end{eqnarray}
so that 
\begin{equation}
  A=\langle\psi_d|\mu_d|\chi_0\rangle + \int \langle\psi_{\epsilon}|\mu_{\epsilon}|\chi_0\rangle d\epsilon
\end{equation}
or after substituting for the wavefunction components
{\small
\begin{eqnarray}\nonumber
  A & = &  \langle\nu_f|V_{d\epsilon_f} \left[E-H_d-F\right]^{-1}  \mu_{d}|\chi_0\rangle 
\\ \nonumber
    && +  \langle\nu_f|\mu_{\epsilon_f}|\chi_0\rangle 
\\ \label{eq:Afinal}
    && +  \langle\nu_f|V_{d\epsilon_f} \left[E-H_d-F\right]^{-1} M(E-h_0)|\chi_0\rangle,
\end{eqnarray}
}
where in analogy with $F(\varepsilon)$ we have defined 
\begin{equation}\label{eq:Mint}
  M(\varepsilon) 
= 
  \int_0^\infty V_{d\epsilon}(R)[\varepsilon-\epsilon+i\eta]^{-1} \mu_{\epsilon}(R') d\epsilon.
\end{equation}
This quantity can be interpreted as the transition amplitude through dipole transition to the continuum 
state $|\epsilon\rangle$ from which the electron is captured to the metastable anion state 
$|\phi_1\rangle$. The three terms in the photodetachment amplitude (\ref{eq:Afinal}) have the 
following interpretation. The most simple is the second term 
$\langle\nu_f|\mu_{\epsilon_f}|\chi_0\rangle$ 
which gives the amplitude for the direct dipole photodetachment from the ground state to 
the bacground continuum 
\begin{equation}\label{eq:Adir}
    \gamma + M^- \to M(\nu_f) + e^-. 
\end{equation}
The first term 
$\langle\nu_f|V_{d\epsilon_f} \left[E-H_d-F\right]^{-1}  \mu_{d}|\chi_0\rangle$
is little bit more complicated as it describes the dipole transition to the discrete 
state $|\phi_1\rangle$ followed by an autodetachment to the neutral molecule and electron 
\begin{equation}\label{eq:A2step}
    \gamma + M^- \to \left(M^-\right)^*\to M(\nu_f) + e^-. 
\end{equation}
The last term describes a three step process of dipole transition to intermediate continuum 
state from which electron is captured to $|\phi_1\rangle$ in second step followed by 
the third step of an autodetachment 
\begin{equation}\label{eq:A3step}
    \gamma + M^- \to e^- + M(\nu) \to \left(M^-\right)^*\to M(\nu_f) + e^-. 
\end{equation}
It is difficult to estimate the relative importance of different mechanisms. We will discuss 
them on a simple model in the next section. Before that we notice that from the calculation 
point of view it is possible to merge the last to process into one expression and 
write the amplitude as the sum of direct and resonance processes 
\begin{equation}
  A =  \langle\nu_f|\mu_{\epsilon_f}|\chi_0\rangle 
    +  \langle\nu_f|V_{d\epsilon_f} |\tilde{\psi}\rangle,
\end{equation}
where the auxilary wavefunction $|\tilde{\psi}\rangle$ is obtained by solution of the equation 
\begin{equation}
  \left[E-T_N-V_1(R)-F\right]|\tilde{\psi}\rangle 
=
  \left[\mu_d+M(E)\right]|\chi_0\rangle.
\end{equation}

\subsection{Photodissociation of the anion}
If the energy of the photon is small enough the process of the dissociative electron detachment 
\begin{equation}
  \gamma + M^- \to e^- + A + B, 
\end{equation}
where $A$ and $B$ are fragments of the neutral molecule $M\equiv AB$ is forbidden.
The potentials sketched in Fig.~\ref{fig:SchemPot} allow for the resonance dissociation 
of the anion 
\begin{equation}
  \gamma + M^- \to \left(M^-\right)^* \to  A^- + B.
\end{equation}
This process is contained in the amplitude of the wave-function $\tilde{\psi}(R)$ 
for $R\to\infty$ or it can be alternatively formulated using the solution with the 
outgoing boudary condition in the potential $V_1(R)$.



%

\section{Test model calculation}

In this part we would like to test the proposed approach on a simple model of electron 
detachment from diatomic anion. The goal of this calculation is not  quantitative description 
of the photodetachment cross sections for any specific molecule, but the parameters of the model 
are on the qualitative level inspired by lithium hydride anion \cite{CDH2018,Ms_Dvorak}.
The model of the photodetachment as described above is determined by knowledge of the 
potential of the neutral molecule $W(R)$, the potential curve of the ground anion state
$V_0(R)$, the excited anion state $V_1(R)$ (the discrete state), the 
discrete-state-continuum coupling function $V_{d\epsilon}(R)$ 
and the dipole moment transition elements 
$\mu_d(R)$ and $\mu_{\epsilon}(R)$. The discrete-state-continuum coupling function $V_{d\epsilon}(R)$
depends on the continuum cahnnel index (for example angular momentum), but in this simple model 
we assume that there is one dominant channel and neglect this dependence. Similarly the dipole 
transition moment is vector quantity, but we assume the fixed polarization with respect 
to molecular axis and treat is as a single number. These details would be important in comparison with 
specific experimental data, but they are not important in this qualitative discussion.

\subsection{Qualitative model for diatomic molecule}
\begin{figure}[th]
\centerline{\includegraphics[width=0.49\textwidth]{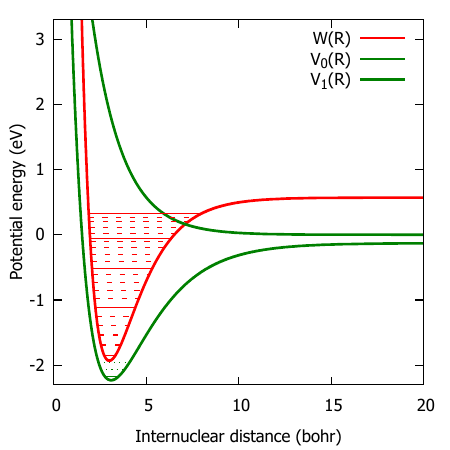}}
\caption{\label{fig:Potentials}%
         Potential energy curves in our model. The neutral molecule is in red and anion
         states in green. The vibrational levels are also shown (every fifth with solid line).}
\end{figure}

The potential energy curves $W(R)$ and $V_0(R)$ for the ground states of the 
neutral molecule and the anion can be calculated by well established mehtods of quantum chemistry.
For LiH molecule high-level calculation is available \cite{GL2006}. We also calculated 
the potential energy curves (see Fig.~1 in \cite{CDH2018}). Here, for the model calculation 
we use simply Morse potential, qulitatively similar to the LiH case. The exact 
form is as follows (numerical values are for the energies in eV and distances 
in Bohr)
\begin{eqnarray}
  W(R)   &=&  d e^{-2a(R-R_0)} - 2d e^{-a(R-R_0)} + b, \\
  V_0(R) &=&  d_0 e^{-2a_0*(R-R_g)} - 2d_0 e^{-a_0*(R-R_g)}+b_0, \\ \label{eq:V1}
  V_1(R) &=&  d_1 e^{-a_1*(R-R_0)},
\end{eqnarray}
where $R_0=3.0$, $a=0.6$, $b=0.57$, $d=2.5$, $R_g=3.1$, $a_0=0.45$, $b_0=-0.13$, $d_0=2.1$, $a_1=0.6$ 
and $d_1=1.9$.

The discrete state potential $V_1(R)$ and its coupling to the continuum $V_{d\epsilon}(r)$
is often extracted by fitting the fixed-nuclei scattering phaseshifts. This is based
on the solution of the fixed-R scattering problem with electronic hamiltonian parametrized 
as 
\begin{eqnarray}\nonumber
  \mathcal{H}_{el} &=&
    |\phi_d\rangle V_{d} \langle\phi_d| +
    \int |\epsilon\rangle \left\{W+\epsilon\right\} \langle\epsilon| d\epsilon \\ \label{eq:HelR}
&& +
    \int \left\{
      |\phi_d\rangle V_{d\epsilon}\langle\epsilon| + |\epsilon\rangle V_{d\epsilon}\langle\phi_d|
         \right\}d\epsilon.
\end{eqnarray}
The solution of the scattering problem $|\Psi(E)\rangle$ at fixed molecular geometry $R$ can be expanded in 
the basis $|\phi_d\rangle$, $|\epsilon\rangle$ in similar way like (\ref{eq:PsiExp})
\begin{equation} \label{eq:fixed_Psi(E)}
  |\Psi(E))\rangle = 
    \psi_{d} |\phi_d\rangle + \int \psi_{\epsilon} |\epsilon\rangle d\epsilon.
\end{equation}
Now $\psi_d$ and $\psi_{\epsilon}$ are numbers (dependent on $R$) not wavefunctions
in vibrational space. It is easy to solve the scattering problem with hamiltonian (\ref{eq:HelR})
and to find the components
\begin{eqnarray} \label{eq:psi_d_comp}
  \psi_d &=& \left[E-V_d-F(E-W)\right]^{-1} V_{d\epsilon_i},
\\ \label{eq:psi_e_comp}
  \psi_{\epsilon} &=& \delta(\epsilon-\epsilon_i) + 
  (E-\epsilon-W)^{-1}V_{d\epsilon}\psi_d.
\end{eqnarray}
\begin{figure}[th]
\centerline{\includegraphics[width=0.25\textwidth]{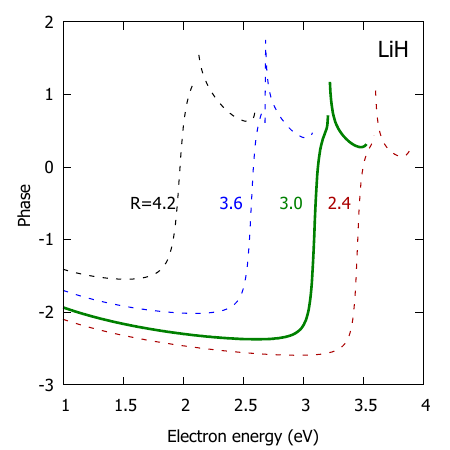}%
\hspace*{-0.1mm}\includegraphics[width=0.25\textwidth]{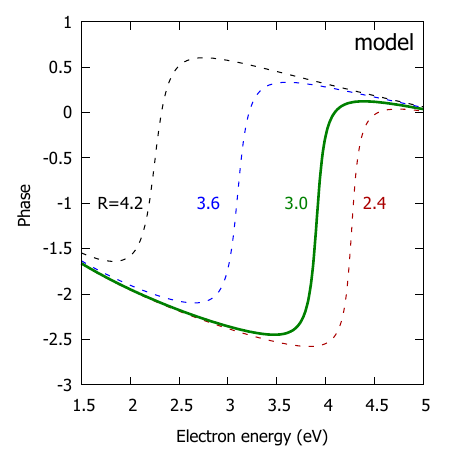}}
\caption{\label{fig:Phase}%
         Fixed-nuclei scattering phaseshifts for few internuclear separations $R$ 
         (labeled in the graphs). Figure in the left shows the results of R-matrix 
         scattering calculations for LiH+e and the right panel is the result 
         from our model.}
\end{figure}
%
The scattering phaseshift for the fixed-nuclei problem then reads\cite{D1991}
\begin{equation}\label{eq:delta}
  \delta=-\arctan\left(
    \frac{\Gamma(\epsilon,R)/2}%
    {\epsilon-V_d(R)-W(R)-\Delta(\epsilon,R)}
  \right),
\end{equation}
with $\Delta$ and $\Gamma$ derived from the real and imaginary part of the fixed-nuclei 
version $F[E-W(R)]$
\begin{equation}\label{eq:Fint}
  F(\varepsilon) = \Delta - {\textstyle\frac{i}{2}\Gamma} 
    = \int V_{d\epsilon}(R)[\varepsilon-\epsilon+i\eta]^{-1} V_{d\epsilon}(R) d\epsilon.
\end{equation}
of the nonlocal level-shift operator $F(E-h_0)$ in (\ref{eq:Fnonloc}). 
Notice that the operator-valued argument $E-h_0$ changed into electron 
energy $\epsilon=E-W(R)$ relative to the scattering threshold.
The fitting of 
the formula (\ref{eq:delta}) to ab initio scattering data for eigenpahses 
is usually used to obtain $V_d(R)$ and $V_{d\epsilon}$. Here we just choose 
the model functions by hand so that the resulting phaseshift (\ref{eq:delta})
is in visual qualitative accordance with the data for LiH molecule \cite{CDH2018}
as demonstrated in Fig.~\ref{fig:Phase}. 

The model function $V_d(R)=V_1(R)$ of Eq.~(\ref{eq:V1}) was choosen to obtain 
the data in the figure and the separable form of discrete-state-continuum coupling
\begin{eqnarray}
  V_{d\epsilon}(R) &=& g(R)f(e), 
\\
  g(R) &=& \left[ 1+e^{0.75(R-6}\right]^{-1},
\\
  \gamma(\epsilon) &\equiv& 2\pi f(\epsilon)^2 
   = A_{\gamma} [\epsilon/B_{\gamma}]^{\alpha} e^{-\epsilon/B_{\gamma}},
\end{eqnarray}
with constants $A_{\gamma}=1$eV, $B_{\gamma}=2$eV, $\alpha=0.2$ was used. 
This form is inspired by earlier studies of electron-molecule collisions
\cite{D1991,Kniha_2012}. It is convenient that the integral transform 
(\ref{eq:Fint}) can be calculated analytically for this form.

\begin{figure}[t]
\centerline{\includegraphics[width=0.25\textwidth]{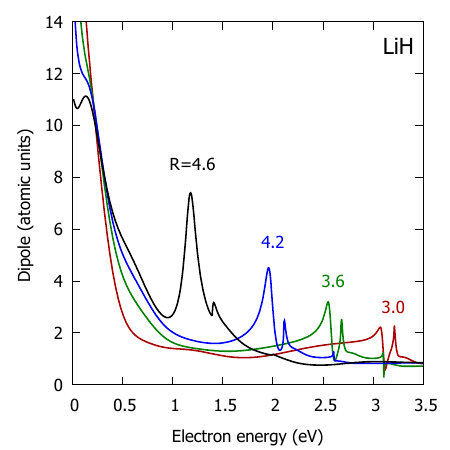}%
\hspace*{-0.1mm}\includegraphics[width=0.25\textwidth]{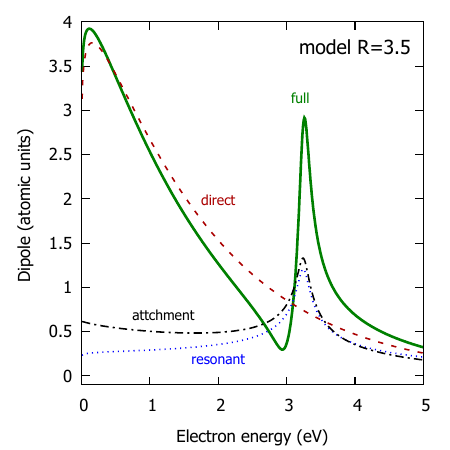}}
\caption{\label{fig:Dip}%
         Transition dipole matrix element from the ground state anion to continuum 
         for fixed-nuclei electronic problem. The internuclear separations $R$
         are marked in the graphs. The left panel show the results of R-matrix
         calculation for LiH molecule, the right panel is calculated from our model, 
         with contributions of three different terms marked separately.}
\end{figure}

The last ingredient of the model is the knowledge of the transition 
dipole moment $\mu_d$ to the discrete state and transition dipole 
moment function $\mu_{\epsilon}$ for each internuclear separation 
$R$. To do so we were again guided by the fixed-nuclei scattering 
calculation \cite{CDH2018,Ms_Dvorak} for LiH. The calculated moment 
function 
$$|\mu|^2=|\langle\chi_0|D|\Psi(E))\rangle|^2$$
is shown in Fig.~(\ref{fig:Dip}) in the left panel%
\footnote{Note that the dipole operator is vector quantity. We show only 
component along the molecular axis. We also include only the lowest partial 
wave in continuum to obtain simple picture for this qualitative model.}, 
but we have to keep in mind that the function $|\Psi(E)\rangle$
has both discrete-state and continuum components (\ref{eq:psi_d_comp}), 
(\ref{eq:psi_e_comp}). Substituting these in (\ref{eq:fixed_Psi(E)}) we 
get relation to $\mu_d$ and $\mu_{\epsilon}$
{\small
\begin{equation}\label{eq:AfixedR}
  \mu  =  \mu_{\epsilon} 
    + \frac{\mu_{d}V_{d\epsilon}}{\epsilon-V_d-F(\epsilon)}  
    + \frac{M(\epsilon)V_{d\epsilon}}{\epsilon-V_d-F(\epsilon)}.
\end{equation}
}
Notice that this expression is fixed-nuclei version of (\ref{eq:A})
and the terms thus have similar interpretation: {\em direct} transition 
dipole to background continuum $\mu_{\epsilon}$ and the next two 
terms are the {\em resonant} contribution due to transition 
to the discrete state and subsequent autodetachment and 
term due to {\em attachment} to the discrete state from the background
continuum. 

We find that this function within our model (see Fig.~\ref{fig:Dip} right panel)
is in reasonable qualitative agreement with the calculation for LiH molecule. 
The three individual contributions are also shown in the figure but the full 
result is not direct sum of the individual contributions, because the complex
phase has to be taken into account. The model functions producing the figure 
are 
\begin{eqnarray}
    \mu_d &=& 0.1 + 0.1 i, 
\\
    2\pi\mu_{\epsilon}^2 &=& A_{\mu}[\epsilon/B_{\mu}]^{\alpha} e^{-\epsilon/B_{\mu}},
\end{eqnarray}
with $A_{\mu}=150$a.u. and $B_{\mu}=0.8$eV.
This form of the functions allows for calculation of the integral transform
in the definition (\ref{eq:Mint}) of function $M(\epsilon)$ analytically in the 
same way as for function $F(\epsilon)$.

\subsection{Notes on numerical treatment}

To calculate the full photodetachment amplitude (\ref{eq:Afinal}) we need to be able 
to apply the operator 
$$|\Psi\rangle=\left[E-H_d-F\right]^{-1}|\Phi\rangle.$$
This is equivalent to solving the equation 
$$\left[E-H_d-F\right]^{-1}|\Psi\rangle = |\Phi\rangle.$$
There are well developed techniques to perform this task in the treatment of the inelastic
electron-molecule collision \cite{D1991,Kniha_2012} and we applied our numerical codes
to perform this task. The operator $F(E-h_0)$ needed there is evaluated by expansion 
into neutral molecule vibrational basis $h_0|\nu\rangle=E_{\nu}|\nu\rangle$ which
converts it to evaluation of the fixed-nuclei quantity $F(\epsilon)$ calculated 
by the integral transform 
\begin{equation}
    F(E-h_0) = \sum_{\nu} |\nu\rangle F(E-E_{\nu})\langle\nu|.
\end{equation}    
The same method can be used to calculate also the operator $M(E-h_0)$.

\begin{figure}[th]
\centerline{\includegraphics[width=0.49\textwidth]{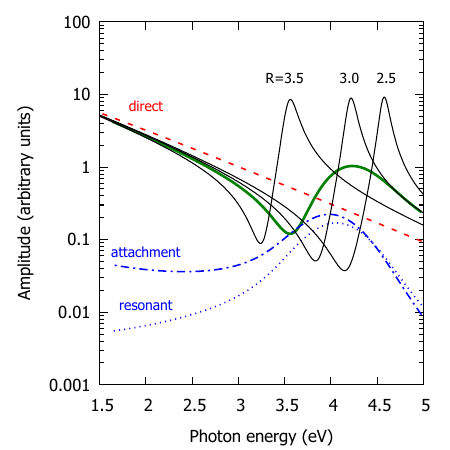}}
\caption{\label{fig:Edep}%
         Dependence of the photodetachment amplitude on the photon energy (full green line).
         The results of full calculation is shown with contribution of different 
         mechanisms also marked separately. Black lines show fixed nuclei amplitudes 
         for three different values of $R$ for comparison.}
\end{figure}

\subsection{Discussion of the results}
\begin{figure}[th]
\centerline{\includegraphics[width=0.25\textwidth]{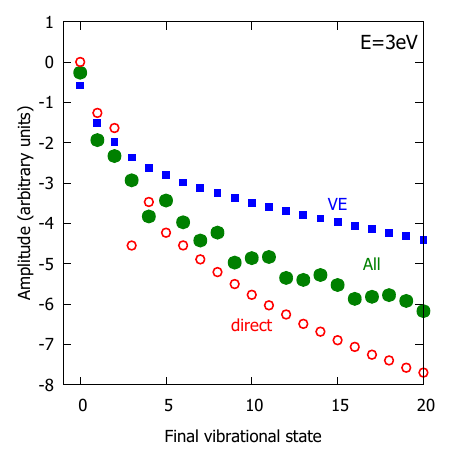}%
\hspace*{-0.1mm}\includegraphics[width=0.25\textwidth]{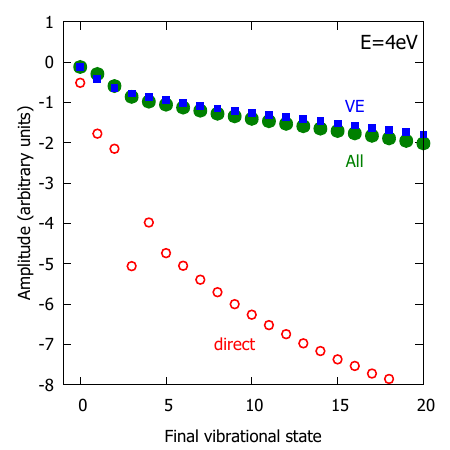}}
\caption{\label{fig:Vdep}%
         Amplitude for photodetachment to different final vibrational states of the neutral molecule.
         Two different photon energies are included $E=3eV$ (off-resonance, left) and 
         $E=4eV$ (in resonance, right). Different symbols show full calculation (full circles)
         and direct photodetachment to background continuum (empty circles). Vibrational excitation 
         cross sections of the neutral molecule by electron scattering are shown for comparison 
         (squares).}
\end{figure}

The calculated results are shown in Fig.~(\ref{fig:Edep}) and (\ref{fig:Vdep}).
In the first of the two figures the full amplitude $|A|^2$ for the final ground vibrational state 
of the neutral molecule formed after the detachment is plotted together with the 
three individual contributions. We see that resonance and attachment contributions 
have similar shape and are important only close to the resonance energy 4eV. 
We also show the fixed nuclei amplitude $|\mu|^2$ for three internuclear 
separations $R$. Observe that the full amplitude is given by smearing of the
fixed nuclei amplitude over the initial vibrational wavefunction $\chi_0(R)$.

In Fig.~(\ref{fig:Vdep}) we show the amplitude for two energies $E=3$eV
and $E=4$eV and for first 20 vibrational states of the final neutral molecule. 
For the energy 3eV, which is below the resonance, the amplitude decreases very 
fast with the vibrational quantum number. On the other hand the resonance 
energy $E=4$eV alows for creation of highly vibrationally excited energies 
which much larger probability. We also see that the background contribution 
continues to decay fast and the resonance and attachment contributions dominate 
for high vibrational quantum numbers.

\section{Conclusions}

We derived theory for calculation of the electron photodetachment from molecular anions 
in resonance condition using the discrete-state-in-continuum model in very similar 
way like in description of inelastic electron-molecule collisions. The techniques 
developed for the numerical treatment of the electron-molecule collisions can therefore 
be directly applied also for resonance photodetachment. We also expect that similar phenomena 
as the ones studied there (threshold peaks, Wigner cusps, boomerang oscillations) can have 
their analogs in photodetachment physics. 

\begin{acknowledgments}
 I thank members of our group Karel Houfek, Jakub Benda, P\v{r}emysl Koloren\v{c} for discussion 
 on the subject during our seminars and especially Zden\v{e}k Ma\v{s}\'{\i}n for encouraging me 
 to finish this work. I also acknowledge the work of my student Ji\v{r}\'{\i} Trnka on the fitting 
 the LiH model.
\end{acknowledgments}

\bibliography{Refs}

\end{document}